\documentclass[aps,prb,reprint,superscriptaddress,amsmath,amssymb,twocolumn,amsfonts,floatfix,longbibliography]{revtex4-2}
\usepackage[colorlinks=true,linktoc=page,linkcolor=blue,citecolor=magenta,urlcolor=electricpurple]{hyperref}
\usepackage{graphicx}
\usepackage{epstopdf}
\usepackage{amsbsy}
\usepackage{mathtools}
\usepackage{gensymb}
\usepackage[dvipsnames]{xcolor}
\usepackage{float}
\definecolor{deepfuchsia}{rgb}{0.76,0.33,0.76}
\definecolor{electricpurple}{rgb}{0.75,0.0,1.0}
\usepackage{orcidlink}

\newcommand{\beq}{\begin{equation}}
\newcommand{\eeq}{\end{equation}}
\newcommand{\bea}{\begin{eqnarray}}
\newcommand{\eea}{\end{eqnarray}}

\newcommand{\fig}[1]{Fig.~\ref{#1}}

\DeclarePairedDelimiterX\braket[2]{\langle}{\rangle}{#1 \delimsize\vert #2}

\begin{document}

\author{Ritu Gupta}
\email[]{ritu.gupta@iitrpr.ac.in}
\affiliation{Department of Physics, Indian Institute of Technology, Ropar 140001, India}
\author{Reena}
\affiliation{Department of Physics, Indian Institute of Technology, Ropar 140001, India}
\author{Dibyendu Samanta}
\affiliation{Department of Physical Sciences, Indian Institute of Technology, Kanpur, India.}
\author{Jaskaran Singh}
\affiliation{Department of Physical Sciences, Indian Institute of Science Education and Research (IISER) Mohali, Knowledge City, Sector 81, Mohali 140306, India.}
\author{Thomas J. Hicken}
\affiliation{PSI Center for Neutron and Muon Sciences, 5232 Villigen, PSI, Switzerland}
\author{Hubertus Luetkens}
\affiliation{PSI Center for Neutron and Muon Sciences, 5232 Villigen, PSI, Switzerland}
\author{Jonas Krieger}
\affiliation{PSI Center for Neutron and Muon Sciences, 5232 Villigen, PSI, Switzerland}
\author{{Sudeep Kumar Ghosh}\,\orcidlink{0000-0002-3646-0629}}
\email[]{skghosh@iitk.ac.in}
\affiliation{Department of Physics, Indian Institute of Technology, Kanpur 208016, India}

\author{Yogesh Singh}
\email[]{yogesh@iisermohali.ac.in}
\affiliation{Department of Physical Sciences, Indian Institute of Science Education and Research (IISER) Mohali, Knowledge City, Sector 81, Mohali 140306, India.}

\date{\today}

\title{Time-Reversal Symmetry Breaking in the Centrosymmetric Nonsymmorphic Superconductor RuB$_2$}

\begin{abstract}
The centrosymmetric nonsymmorphic $A$B$_2$ ($A=$ Ru, Os) borides exhibit several anomalous superconducting properties, yet the microscopic nature of their superconducting ground state remains unresolved. Here we investigate RuB$_2$ using zero- and transverse-field muon spin relaxation ($\mu$SR) measurements, first-principles electronic-structure calculations, and Ginzburg--Landau symmetry analysis. Zero-field $\mu$SR reveals spontaneous internal magnetic fields below $T_c\approx1.6$ K, providing direct evidence for time-reversal symmetry breaking, while transverse-field measurements establish a fully gapped superconducting state with a gap magnitude close to the BCS weak-coupling limit. Density-functional calculations further show that RuB$_2$ is a multiband $\mathbb{Z}_2$ topological metal with appreciable spin--orbit coupling. Guided by the experimental observations and the symmetry of the orthorhombic crystal structure, our Ginzburg--Landau analysis suggests that an internally antisymmetric nonunitary spin-triplet state is a promising candidate for the superconducting order parameter, consistent with both the observed time-reversal symmetry breaking and the fully gapped superconducting state. These results establish RuB$_2$ as a promising platform for investigating unconventional superconductivity arising from the interplay of crystal symmetry, multiband electronic structure, and spin--orbit coupling.
\end{abstract}

\maketitle 

\section{Introduction}
Time-reversal symmetry breaking (TRSB) in superconductors is widely regarded as one of the clearest manifestations of unconventional Cooper pairing because it signals a superconducting order parameter with a complex internal phase structure beyond conventional BCS theory \cite{Sato,Ghosh2020a,Qi,Maeno2001}. Over the past decade, zero-field muon spin relaxation ($\mu$SR) measurements have established TRSB in a relatively small but steadily growing family of superconductors, including strongly correlated materials, noncentrosymmetric compounds, and systems with nontrivial electronic topology \cite{Shang2018,Hillier2012,Ghosh2020b,Biswas2021chiral,Shang2020,Ghosh2022Dirac,Shang2022Weyl,Sajilesh2025time,Kataria2026}. These discoveries have considerably broadened the landscape of unconventional superconductivity, yet a unified microscopic understanding remains elusive because TRSB has emerged in materials with widely different crystal symmetries, electronic structures, and interaction strengths \cite{Ghosh2020a,samanta2026, Ghosh2021time, bhattacharyya2018unconventional,singh2020time,mayoh2021evidence,barker2015unconventional,bhattacharyya2022nodeless,anand2023time, aarti2024time}. Identifying new material platforms that exhibit TRSB despite the apparent absence of conventional ingredients such as strong electronic correlations or inversion-symmetry breaking is therefore essential for uncovering alternative routes to unconventional superconductivity.

Recently, superconductors crystallizing in nonsymmorphic crystal structures have attracted growing attention because glide and screw symmetries can enforce symmetry-protected band degeneracies, hourglass dispersions, and nontrivial topological electronic states that fundamentally influence superconductivity \cite{Mardanya2024,Ghosh2021time,Badger2022,srivastava2026,Yadav2024,Ghosh2023zrossi,Meena2025nonsym,Meena2025}. Several nonsymmorphic superconductors have consequently emerged as promising candidates for unconventional and topological superconductivity, while a growing subset also exhibits spontaneous TRSB below $T_c$, including LaNiGa$_2$, La(Pt,Ni)(Si,Ge), Zr$_3$Ir, HfRhGe, LaPt$_3$P, and the recently discovered type-I TRSB superconductor YbSb$_2$ \cite{Hillier2012,Ghosh2020b,Badger2022,Shang2022Weyl,Shang2020,Ghosh2022Dirac,Biswas2021chiral,Sajilesh2025time,Kataria2026}. Together, these discoveries suggest that crystal symmetry, spin--orbit coupling, and multiband electronic structure may provide additional routes to unconventional superconductivity beyond the conventional paradigms based on strong correlations or broken inversion symmetry.

The centrosymmetric nonsymmorphic $A$B$_2$ ($A=$ Ru, Os) borides provide an attractive setting to explore these ideas. Both compounds crystallize in the orthorhombic space group $Pmmn$ (No.~59) and exhibit several unusual superconducting characteristics \cite{Jaskaran,Singh2007,Singh2010}. These include an upward curvature of $H_c(T)$, a reduced specific-heat jump at $T_c$, non-BCS temperature dependence of the magnetic penetration depth, unusually small Ginzburg--Landau parameters ($\kappa \sim 1$), and signatures of first-order superconducting transitions in magnetic field \cite{Jaskaran,Singh2007,Singh2010}. Both OsB$_2$ and RuB$_2$ possess similar multiband Fermi surfaces, with RuB$_2$ differing primarily by an additional small Fermi pocket \cite{Singh2010,Hebbache2009,Jaskaran}. These observations initially motivated proposals of two-gap superconductivity, supported by penetration-depth measurements in OsB$_2$ and heat-capacity analysis in RuB$_2$ \cite{Singh2010,Jaskaran}. Subsequently, the calculated gap spectrum instead supported a single but strongly anisotropic $s$-wave gap in OsB$_2$ \cite{Bekaert2016}, while electronic-structure calculations identified OsB$_2$ as a $\mathbb{Z}_2$ topological metal with spin-polarized surface states \cite{l3bd-c4my}. In contrast, despite its closely related electronic structure, appreciable spin--orbit coupling, and multiband character, the microscopic superconducting state of RuB$_2$ has remained largely unexplored.

In this work, we investigate the superconducting state of RuB$_2$ by combining zero- and transverse-field $\mu$SR measurements with first-principles electronic-structure calculations and Ginzburg--Landau symmetry analysis. The $\mu$SR measurements establish RuB$_2$ as a type-II superconductor with a fully developed superconducting gap whose temperature dependence is well described by the BCS weak-coupling limit, while zero-field measurements reveal spontaneous internal magnetic fields below $T_c$, providing direct evidence for TRSB. Our electronic-structure calculations further show that RuB$_2$ is a multiband $\mathbb{Z}_2$ topological metal with appreciable spin--orbit coupling. Taken together, these experimental and theoretical results place important constraints on the superconducting order parameter and suggest that multiband or interorbital pairing degrees of freedom may play an important role in the superconducting state. More broadly, RuB$_2$ provides a promising platform for investigating unconventional superconductivity arising from the interplay of crystal symmetry, multiband electronic structure, and spin--orbit coupling.

\section{Synthesis and Characterization of Bulk Superconductivity}
Polycrystalline samples of RuB$_2$ were synthesized by arc-melting stoichiometric ratios of Ru (5N, Alfa Aesar) and B (6N, Alfa Aesar) as described in detail previously \cite{Singh2007,Jaskaran}. Powder x-ray diffraction shown in Fig.~SM1(a) confirmed that the synthesized material is single phase and a refinement of the powder pattern gave lattice parameters which match well with the reported values \cite{Singh2007,Jaskaran}.The electrical resistivity $\rho$ versus temperature $T$ data for RuB$_2$ in zero applied magnetic field are shown in Fig.~SM1(b) in supplementary between $T = 0.5$~K and $7$~K\@.  The Fig.~SM1(c) shows the heat capacity $C(T)$ data below $T = 3$~K measured in zero field and in an applied magnetic field of 150~Oe.  The sharp drop to zero resistance and the sharp anomaly in $C(T)$ at $T_c \approx 1.5$~K signals the onset of bulk superconductivity in RuB$_2$ consistent with previous reports \cite{Jaskaran}.

\section{\text{$\mu$SR} Measurements}
Muon spin relaxation/rotation ($\mu$SR) measurements were carried out at the Paul Scherrer Institute (PSI), Villigen, Switzerland, using the FLAME (FLexible Advanced MuSR Environment) spectrometer. Experiments were performed in both zero-field (ZF) and transverse-field (TF) configurations to investigate the magnetic and superconducting properties of the sample over the temperature range 0.01--2~K.
In the TF configuration, a magnetic field is applied perpendicular to the initial muon spin polarization. The sample is cooled below its superconducting transition temperature to the base temperature of the spectrometer, and data is collected during the warming cycle. In contrast, ZF measurements were performed in a statically compensating ambient field to a level 5~$\mu$T at the sample position.
The $\mu$SR data were analyzed using the \texttt{MUSRFIT} software ~\cite{suter2012musrfit}.

\subsection{Transverse-field measurements}
\begin{figure*}[!t]
    \centering
    \includegraphics[width=1.0\linewidth]{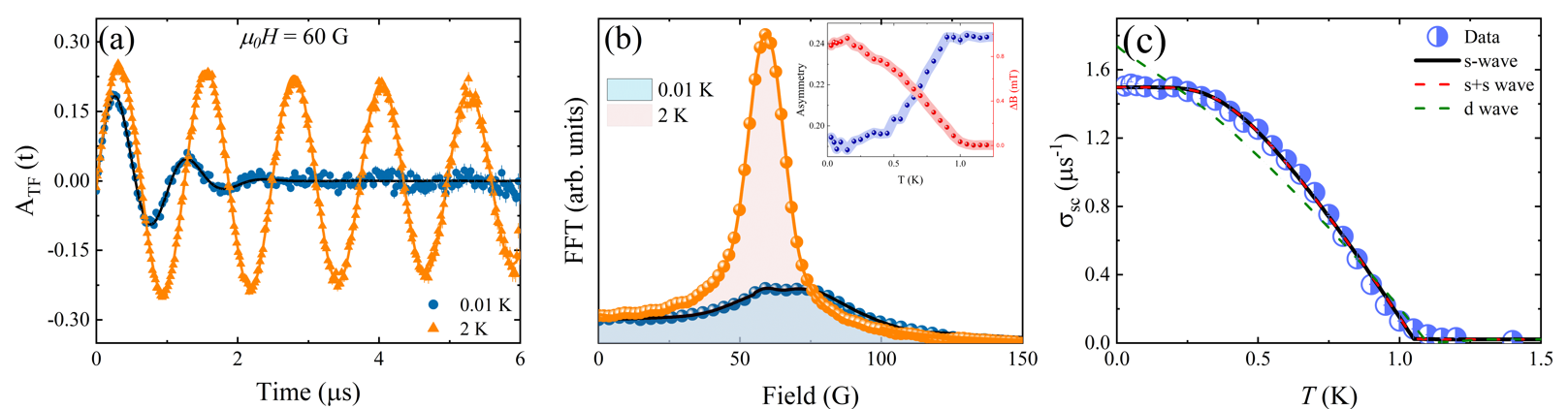}
    \caption{\textbf{TF-$\mu$SR Response of RuB$_2$:} (a) TF-$\mu$SR time spectra of RuB$_2$ measured above $T_c$ (orange symbols) and below $T_c$ (blue symbols) in applied field of 60~G. The solid lines represent fits to the data as described in Supplementary Material (SM). (b) Corresponding Fourier transforms of the TF-$\mu$SR spectra above and below $T_c$. A sharp peak is observed around 60~G above $T_c$, while below $T_c$ the peak broadens and exhibits a positive shift, indicating an increase in the internal magnetic field sensed by the muons in the superconducting state. The inset shows the temperature dependence of the internal field shift, $\Delta B = B_{\mathrm{int}} - B_{\mathrm{ext}}$, and the asymmetry associated with the sample signal. (c) Temperature dependence of the superconducting depolarization rate, $\sigma_{\mathrm{sc}}(T)$. The solid red and green lines represent fits using single-gap $s$-wave and multi-gap $s+s$-wave models, respectively, while the dotted blue line corresponds to a $d$-wave model fit.}
    \label{fig:TF-spectra}
\end{figure*}
Transverse-field (TF) $\mu$SR measurements were performed in an applied field of $\mu_0H_{\mathrm{app}}=60$~G to investigate the superconducting state of $\mathrm{RuB}_2$. The field was applied along the muon-beam direction, with the initial muon-spin polarization oriented at 60$\degree$ to the applied field. The muon-decay positron asymmetry was detected along a direction perpendicular to both the applied field and the initial spin polarization. Representative asymmetry spectra collected above ($2$~K) and below $T_c$ ($0.01$~K) are shown in Fig.~\ref{fig:TF-spectra}(a). Previous studies have suggested that $\mathrm{RuB}_2$ is a type-I superconductor with a critical field of approximately $\mu_0H_c\sim120$~G. For a type-I superconductor with a finite demagnetization factor, the intermediate state is expected when the applied field lies in the range $(1-N)H_c < H_{\mathrm{app}} < H_c$. Considering an average demagnetization factor of $N\approx0.72$ for our cylindrical pellet, the intermediate state is therefore expected for applied fields between approximately 34 and 120~G. In this regime, the sample is expected to separate into superconducting (Meissner) and normal domains. Consequently, the TF-$\mu$SR field distribution is expected to contain three characteristic components: a low-field component associated with muons stopping in the Meissner regions, a component at the applied field arising from the background signal, and a higher-field component, close to the thermodynamic critical field, associated with the normal domains \cite{PhysRevResearch.2.023142, PhysRevB.104.L100508, PhysRevResearch.2.023142, PhysRevB.102.014514}.

At $\mu_0H_{\mathrm{app}}=60$~G, which lies well within the expected intermediate-state regime, we do not observe a distinct high-field component near the reported critical field of 120~G. Moreover, the observation of an almost full oscillating asymmetry suggests that there cannot be a large fraction of low-field regions associated with Meissner regions. These observations do not provide a clear signature of the conventional type-I intermediate state in our sample. Instead, the dominant field distribution is consistent with a vortex-state response characteristic of type-II superconductivity. Nevertheless, the absence of well-resolved intermediate-state features alone does not completely exclude a possible type-I contribution. Sample inhomogeneity, a distribution of local critical fields, deviations from the ideal cylindrical geometry, and the possible coexistence of different superconducting regions could broaden or obscure the expected field components. We therefore conclude that our data do not show a clear and dominant type-I intermediate-state signature, while indicating predominantly type-II-like superconducting behavior.

The spectra above $T_c$ exhibit only a weak relaxation, which can be attributed to static, randomly distributed local magnetic fields associated with the nuclear magnetic moments. Upon entering the superconducting state, the muon-spin depolarization rate increases due to the development of an inhomogeneous internal magnetic-field distribution associated with the superconducting state. The Fast Fourier Transform (FFT) spectra obtained using the maximum-entropy Fourier transform technique are shown in Fig.~\ref{fig:TF-spectra}(b). Interestingly, as highlighted in the inset of Fig.~\ref{fig:TF-spectra}(b), the internal field experienced by the muons exhibits an anomalous paramagnetic shift below $T_c$, rather than the diamagnetic shift conventionally expected for a type-II vortex state. Similar anomalous paramagnetic shifts have been reported in other superconducting systems, including $\mathrm{Ta_2V_{3.1}Si_{0.9}}$~\cite{graham2024microscopic} and $\mathrm{Nb_{0.25}Bi_2Se_3}$~\cite{PhysRevB.102.134514}.

The temperature dependence of the superconducting depolarization rate, $\sigma_{\mathrm{sc}}(T)$, extracted from fits to the TF-$\mu$SR spectra as described in the Supplementary Material, is shown in Fig.~\ref{fig:TF-spectra}(c). The data are equally well described by both a single-gap isotropic $s$-wave model and a two-gap $s+s$-wave model, with a reduced $\chi^2$ value of 0.99 for both fits. The single-gap $s$-wave model yields a gap magnitude of $\Delta_0=0.15(3)$~meV, corresponding to $\Delta_0/k_BT_c=1.63(3)$, which is close to the weak-coupling BCS value of 1.76. In comparison, the $s+s$-wave model gives $\Delta_1/k_BT_c=1.97(6)$ and $\Delta_2/k_BT_c=0.91(8)$, with a weight fraction of approximately 0.86 for the larger-gap component. Previous studies of $\mathrm{RuB}_2$ have reported evidence for multigap superconductivity from specific-heat measurements described using an $s+s$-wave model~\cite{PhysRevB.97.054506}, as well as direct spectroscopic evidence for two superconducting energy gaps~\cite{datta2020spectroscopic}. Our TF-$\mu$SR results are therefore consistent with the presence of multigap superconductivity in $\mathrm{RuB}_2$, although the present $\mu$SR data alone do not uniquely distinguish between the single-gap and multigap models.

In addition to the evolution of the field distribution, we observe a reduction in the total muon-spin asymmetry below $T_c$ [see the inset of Fig.~\ref{fig:TF-spectra}(b)]. Such a reduction can arise from an increased fraction of muons experiencing a broad or strongly inhomogeneous internal-field distribution and may be associated with superconducting regions with different local magnetic environments. In the context of the reported type-I behavior of RuB$_2$, this observation may indicate that a fraction of the sample enters a low-field Meissner-like state. However, the observation of an almost fully oscillating signal prevents us from assigning this feature unambiguously to a type-I intermediate state. Alternatively, the apparent reduction in the initial asymmetry may result from very rapid depolarization in regions with a broad internal-field distribution. A third possibility is that the apparent reduction in the initial asymmetry is a fitting artifact arising from the correlation between the initial asymmetry and the damping parameters when extracting the asymmetry from a strongly damped signal. The combined observations therefore point toward a more complex superconducting state than that expected for a simple conventional type-I or type-II superconductor.
\begin{figure*}
    \centering
    \includegraphics[width=1.0\linewidth]{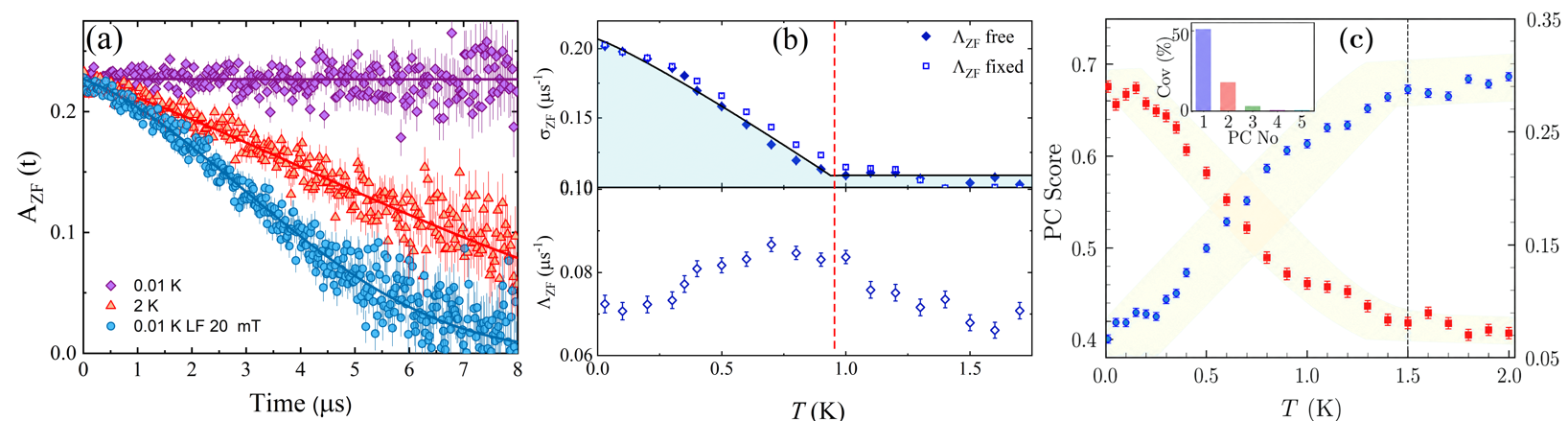}
    \caption{\textbf{Spontaneous Internal Fields in RuB$_2$:} (a) ZF-$\mu$SR spectra of RuB$_2$ measured at 0.01~K (below $T_c$, blue symbols) and 2~K (above $T_c$, red symbols), showing a clear change in the relaxation rate across the superconducting transition. The enhanced relaxation below $T_c$ indicates the emergence of spontaneous internal magnetic fields, suggesting TRSB. The violet symbols represent measurements performed under an applied longitudinal field of 20 mT, where the relaxation is significantly suppressed, indicating the static/quasi-static nature of the internal fields. (b) Temperature dependence of the relaxation rate obtained from fitting the ZF-$\mu$SR data. The relaxation rate increases with decreasing temperature and is well described by order parameter like behavior. (c) Temperature evolution of the first and second principal component scores extracted from a principal component analysis of the $\mu$SR spectra. The inset displays the fraction of total variance captured by each principal component. Shaded regions and dashed lines are included as visual guides to the eye.}
    \label{fig:ZF spectra}
\end{figure*}
An interesting possibility is that the unusual superconducting response of $\mathrm{RuB}_2$ may be related to its multiband nature. Both type-I/type-II behavior near the boundary $\kappa\approx1/\sqrt{2}$ and type-1.5 superconductivity can give rise to non-monotonic vortex-vortex interactions, with competing long-range attractive and short-range repulsive contributions. While type-I/type-II crossover behavior is confined to a narrow region of the $\kappa$--$T$ phase diagram, type-1.5 superconductivity can naturally arise in multiband superconductors when the characteristic length scales associated with different superconducting condensates lead to competing interactions. Given the growing experimental evidence for multigap superconductivity in $\mathrm{RuB}_2$, the present results make it a potentially interesting candidate for type-1.5 superconductivity. However, distinguishing conclusively between a type-I/type-II crossover and a type-1.5 superconducting state requires further measurements, including detailed field-dependent TF-$\mu$SR studies, determination of the relevant length scales, and investigation of the vortex-field distribution over a wider range of applied fields.

\subsection{Zero-field measurements}
Next, ZF-$\mu$SR measurements were performed to probe spontaneous internal magnetic fields associated with time-reversal symmetry breaking (TRSB). The ZF spectra are well described by a Gaussian Kubo--Toyabe function with an additional exponential relaxation, as shown for representative temperatures above and below ($T_c$) in Fig.~\ref{fig:ZF spectra}(a). A detailed analysis of the ZF spectra is provided in the Supplementary Material. Upon entering the superconducting state, a clear enhancement of the Gaussian relaxation rate, ( $\sigma_{\mathrm{ZF}}$), is observed which is associated with static nuclear dipolar fields. The temperature dependence of ($\sigma_{\mathrm{ZF}}$) [Fig.~\ref{fig:ZF spectra}(b)] exhibits a distinct increase below ($T_c$), indicating the emergence of spontaneous internal fields in the superconducting state. The corresponding spontaneous internal field can be estimated from the change in the relaxation rate, ($\Delta\sigma_{\mathrm{ZF}}$), using $B_{\mathrm{spont}}=\sqrt{2}\Delta\sigma_{\mathrm{ZF}}/\gamma_\mu$, yielding ($B_{\mathrm{spont}}\approx0.152~\mathrm{mT}$). This spontaneous field is comparable in magnitude to those reported for several established TRSB superconductors, including Sr$_2$RuO$_4$ \cite{grinenko2021unsplit} and multiband iron-based superconductors \cite{grinenko2020superconductivity}.

No oscillatory signal is observed in the ZF-$\mu$SR spectra, ruling out the presence of sizable static long-range magnetic order within the sensitivity of the measurement. Furthermore, the application of a longitudinal field (LF) of 20~mT fully suppresses the additional relaxation, indicating that the spontaneous internal fields are static or quasistatic on the muon timescale. Although 20~mT is relatively large compared with the weak internal fields associated with TRSB, measurements performed on RuB$_2$ using the DOLLY spectrometer, presented in the Supplementary Material, demonstrate that an LF of only 9~mT is sufficient to fully decouple the muon spins from the spontaneous internal fields below ($T_c$). This further supports the static or quasistatic nature of the spontaneous internal fields. A broad hump in \(\lambda_{\mathrm{ZF}}\) around \(T_c\), as shown in Fig.~\ref{fig:ZF spectra}(b), is also observed, which may be associated with the development of weak spontaneous fields and their possible spatial inhomogeneity. Similar behavior has been reported in other TRS-breaking superconductors, including Sr$_{0.1}$Bi$_2$Se$_3$ \cite{Neha}, Re$_6$Zr \cite{RP}, and Pr$_{1-x}$Ce$_x$Pt$_4$Ge$_{12}$ \cite{Jian}. Also, the decoupling of the additional ZF relaxation under a weak longitudinal field supports an intrinsic origin of the observed relaxation and argues against extrinsic effects such as dilute magnetic impurities.

The observation of spontaneous internal fields, together with the fully gapped multiband superconducting response obtained from TF-$\mu$SR, is particularly noteworthy. In multiband superconductors, TRSB can arise from a nontrivial relative phase between different superconducting condensates. The present results therefore suggest the possibility of a complex multiband superconducting state in RuB$_2$, although the present measurements alone do not uniquely determine the symmetry or relative phase of the superconducting gaps. Taken together, these results provide evidence for spontaneous TRSB in the superconducting state of RuB$_2$ and, alongside the unusual magnetic response observed in TF-$\mu$SR, point toward a potentially unconventional multiband superconducting state.
\section{Electronic band structure and normal state topology}
\begin{figure*}[!htb]
\includegraphics[width=\textwidth]{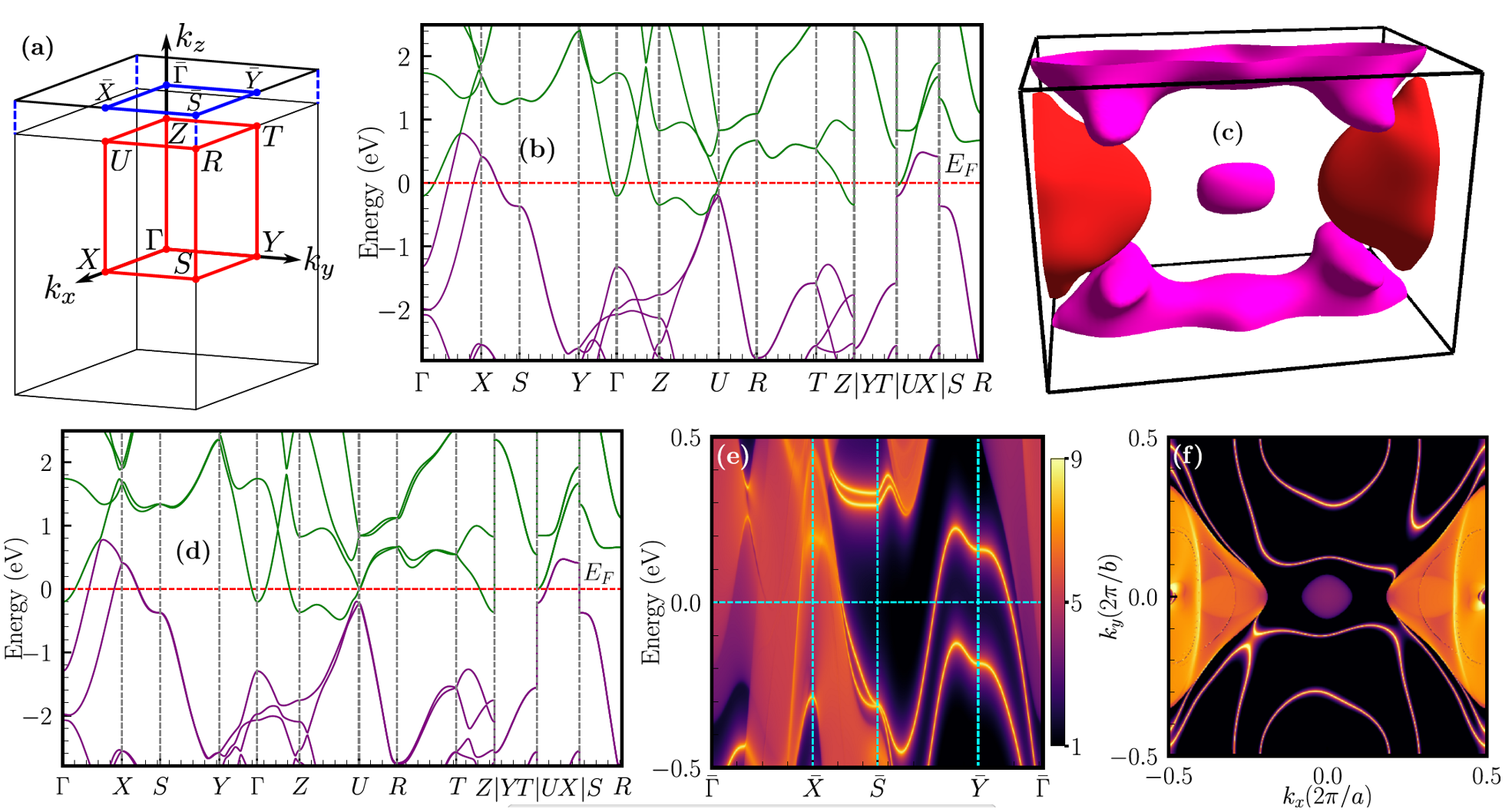} 
\caption{\textbf{Electronic band structure, and band topology of RuB$_2$:} (a) Bulk Brillouin zone corresponding to the primitive orthorhombic lattice, including $(001)$ surface Brillouin zone. (b) Electronic band structure calculated without spin-orbit coupling (SOC). (c) Combined Fermi surface sheets of RuB$_2$, highlighting three Fermi sheets spanning the Brillouin zone. (d) Electronic band structure with SOC included, showing SOC-induced band splitting. (e) Surface spectral function along high-symmetry paths of the $(001)$ surface BZ, revealing multiple topological surface states. (f) Constant-energy surface spectrum in the vicinity of the Fermi level.}         
\label{fig:bands_topology}
\end{figure*}
We performed first-principles density functional theory (DFT) calculations to investigate the electronic band structure of RuB$_2$. The exchange-correlation effects were treated within the generalized gradient approximation (GGA), as implemented in the QUANTUM ESPRESSO package~\cite{Giannozzi_2009,Giannozzi_2017,Giannozzi_2020}. RuB$_2$ crystallizes in an orthorhombic structure with the nonsymmorphic space group Pmmn (No. 59), belonging to the point group $D_{2h}$. The lattice parameters are $a = 2.8657~\mathrm{\AA}$, $b = 4.6457~\mathrm{\AA}$, and $c = 4.0462~\mathrm{\AA}$. The primitive unit cell contains two formula units, comprising two Ru atoms and four B atoms. Structural optimization, including both atomic positions and lattice parameters, was carried out using variable-cell relaxation. A plane-wave kinetic energy cutoff of $80$ Ry was employed throughout the calculations. Brillouin zone integrations were performed using an $18 \times 11 \times 13$ Monkhorst-Pack $k$-point mesh.

\fig{fig:bands_topology}(a) illustrates the bulk Brillouin zone of the primitive orthorhombic lattice. The corresponding electronic dispersion in the absence of spin–orbit coupling (SOC), shown in \fig{fig:bands_topology}(b), reveals three doubly degenerate and strongly dispersive bands crossing the Fermi level. This leads to the formation of both electron- and hole-like pockets, highlighting the intrinsically multiband nature of RuB$_2$. The low-energy electronic states predominantly originate from Ru-$4d$ orbitals, with additional contributions from B-$2p$ states, as confirmed by the projected density of states [see Supplementary Material (SM)]. Upon including SOC, the band degeneracies are lifted, and a continuous direct gap emerges between the valence (purple) and conduction (green) manifolds near the Fermi energy, as illustrated in \fig{fig:bands_topology}(d). Nevertheless, RuB$_2$ remains metallic, as the strongly dispersive bands near the Fermi level continue to sustain sizable electron and hole pockets. Moreover, owing to the centrosymmetric crystal structure, Kramers’ theorem guarantees that all bands remain at least doubly degenerate throughout the Brillouin zone; consequently, SOC induces only modest modifications to the Fermi surfaces. The resulting Fermi surface, shown in \fig{fig:bands_topology}(c) with SOC included, comprises three sheets: two hole pockets centered at the $X$ point and one electron pocket centered at the $Y$ point of the Brillouin zone.

The existence of a continuous direct gap between the valence and conduction bands across the entire Brillouin zone in RuB$_2$ permits a topological classification analogous to that of a fully gapped insulator~\cite{Schoop2015,Nayak2017,Ortiz2020}. To characterize its topology, we evaluate the $\mathbb{Z}_2$ invariant by tracking the evolution of hybrid Wannier charge centers (WCCs) throughout the Brillouin zone~\cite{Yu2011,Soluyanov2011}. The resulting WCC flow yields a $\mathbb{Z}_2$ index of $(0;010)$, thereby establishing RuB$_2$ as a $\mathbb{Z}_2$ topological metal.

To further elucidate the topological nature of RuB$_2$, we compute the surface spectral function by projecting the bulk electronic structure onto the (001) surface. The resulting spectrum reveals multiple topologically nontrivial surface states crossing the Fermi level, including a prominent surface Dirac cone at $\bar{S}$ (approximately $-0.3~\text{eV}$), in agreement with the nontrivial $\mathbb{Z}_2$ invariant, as shown in \fig{fig:bands_topology}(e). The corresponding constant-energy surface spectrum in the vicinity of the Fermi level is presented in \fig{fig:bands_topology}(f). These robust surface states provide a favorable platform for realizing topological superconductivity via proximity coupling to the bulk superconducting condensate.

\section{Discussion and summary}  
The coexistence of time-reversal symmetry breaking and a fully gapped superconducting state places strong constraints on the nature of the superconducting order parameter in RuB$_2$. To examine the symmetry-allowed pairing instabilities, we employ a Ginzburg--Landau symmetry analysis for the orthorhombic point group $D_{2h}$~\cite{Ghosh2020a,Annett1990,sigrist1991}. In the presence of appreciable spin--orbit coupling, all irreducible representations of $D_{2h}$ are one-dimensional, implying that symmetry-allowed single-band superconducting states are nondegenerate and therefore cannot spontaneously break time-reversal symmetry~\cite{sumita2018}. In the weak-SOC limit, where the normal-state symmetry is enlarged to $D_{2h}\otimes SO(3)$, four TRS-breaking superconducting instabilities are symmetry allowed~\cite{Hillier2012,Weng2016}. However, all of these states necessarily possess point nodes, in contrast with the fully gapped superconducting state established by our transverse-field $\mu$SR measurements. Furthermore, although nonsymmorphic crystal symmetries can protect symmetry-enforced nodes along the Brillouin-zone boundary, they do not generate the multicomponent order parameters required for spontaneous TRS breaking~\cite{sumita2018}. These considerations suggest that a conventional single-band pairing description is unlikely to account for the combined observation of broken TRS and a fully gapped superconducting state in RuB$_2$.

Our first-principles calculations instead reveal that multiple Ru-$4d$ orbitals contribute significantly to the electronic states at the Fermi level, resulting in a multiband Fermi surface composed of several closely spaced sheets. Such an electronic structure naturally motivates superconducting states involving interband or interorbital pairing, where additional pairing channels become available beyond a single-band description. In particular, the presence of multiple orbital degrees of freedom enables phase frustration between competing superconducting condensates, providing a natural route to spontaneous TRS breaking while preserving a fully open quasiparticle gap. Moreover, the nearly parallel and closely spaced Fermi-surface sheets found in RuB$_2$ provide a favorable electronic environment for stabilizing interorbital pairing states.

In this context, nonunitary triplet states with internal antisymmetry in orbital space have been proposed as promising candidates for fully gapped TRS-breaking superconductivity in centrosymmetric systems with the same orthorhombic point-group symmetry~\cite{Hillier2012,Weng2016,Ghosh2020b,Badger2022}. Motivated by the multiorbital electronic structure revealed by our first-principles calculations, we consider such an internally antisymmetric nonunitary triplet (INT) pairing state as a plausible candidate for the superconducting ground state of RuB$_2$. Within this scenario, Cooper pairs form between electrons occupying different orbitals on the same lattice site while residing in a spin-triplet channel. The required fermionic antisymmetry is therefore encoded in the orbital sector, allowing a fully gapped superconducting state despite the triplet spin configuration. The pairing potential can be written as $\hat{\Delta}=\hat{\Delta}_{\mathrm{spin}}\otimes\hat{\Delta}_{\mathrm{orb}}$, where $\hat{\Delta}{\mathrm{spin}}=(\mathbf{d} \cdot \boldsymbol{\sigma})i\sigma_y$ and $\hat{\Delta}_{\mathrm{orb}}=i\tau_y$, with $\boldsymbol{\sigma}$ ($\boldsymbol{\tau}$) denoting the Pauli matrices in spin (orbital) space. The triplet pairing is characterized by the $\mathbf{d}$ vector, $\mathbf{d}=\Delta_0\boldsymbol{\eta}$ with $|\boldsymbol{\eta}|^2=1$, while the complex vector $\boldsymbol{\eta}$ gives rise to a nonunitary state through the finite spin polarization $\mathbf{q}=i(\boldsymbol{\eta}\times\boldsymbol{\eta}^{*})\neq0$. Here, $\Delta_0$ denotes a uniform pairing amplitude, consistent with the experimentally observed fully gapped superconducting state in RuB$_2$. Such a pairing state can naturally generate spontaneous internal magnetic fields without requiring either broken inversion symmetry or strong electronic correlations. Although establishing the microscopic pairing interaction lies beyond the scope of the present work, the combined experimental observations, electronic-structure calculations, and symmetry analysis suggest that superconducting states involving multiband or interorbital degrees of freedom provide a natural framework for understanding the superconducting ground state of RuB$_2$.

RuB$_2$ therefore extends the growing family of superconductors in which broken time-reversal symmetry coexists with a fully gapped quasiparticle spectrum, while doing so in a centrosymmetric, weakly correlated, and nonsymmorphic material. Together with its multiband $\mathbb{Z}_2$ topological metallic normal state, these results identify RuB$_2$ as an attractive platform for investigating how crystal symmetry, electronic topology, and multiband pairing cooperate to produce unconventional superconductivity. More broadly, our work highlights that spontaneous TRS breaking need not be restricted to strongly correlated or noncentrosymmetric systems, but may also emerge from the interplay of orbital degrees of freedom, spin--orbit coupling, and symmetry in comparatively simple multiband superconductors. Future studies combining phase-sensitive probes, spectroscopic measurements, and microscopic theoretical modelling will be valuable for establishing the precise pairing mechanism and elucidating the role of topology in the superconducting state.
\\

\emph{Acknowledgments:} 
This work was performed at the Swiss Muon Source
(S$\mu$S) Paul Scherrer Institute, Villigen, Switzerland. We thank the X-ray facility at IISER Mohali. RG acknowledges the Department of Science and Technology (DST), India, for financial support under the INSPIRE grant (Grant No. DST/INSPIRE/04/2022/002178), and the Anusandhan National Research Foundation (ANRF) for support under the PMECRG scheme (Grant No. ANRF/ECRG/2024/003930/PMS) and the Partnerships for Accelerated Innovation and Research (PAIR) scheme (Project No. ANRF/PAIR/2025/000006/PAIR-A). R. acknowledges financial support from the Council of Scientific and Industrial Research (CSIR), India, through fellowship. JS acknowledges UGC-CSIR India for a fellowship. SKG acknowledges financial support from Anusandhan National Research Foundation (ANRF) erstwhile Science and Engineering Research Board (SERB), Government of India via the Startup Research Grant: SRG/2023/000934 and from IIT Kanpur via the Initiation Grant (IITK/PHY/2022116). SKG and DS utilized the \textit{Andromeda} server at IIT Kanpur for numerical calculations.\\
\onecolumngrid
\clearpage

\clearpage

\onecolumngrid
\noindent
\begin{center}
    {\Large\bfseries
    Supplementary Materials for\\[0.3em]
    ``Time-Reversal Symmetry Breaking in the\\
    Centrosymmetric Nonsymmorphic Superconductor RuB$_2$''
    }
\end{center}

\vspace{1em}
\twocolumngrid
\setcounter{figure}{0}
\renewcommand{\thefigure}{S\arabic{figure}}
\section{Basic Characterizations}
\begin{figure*}[t]   
\includegraphics[width= 7 in]{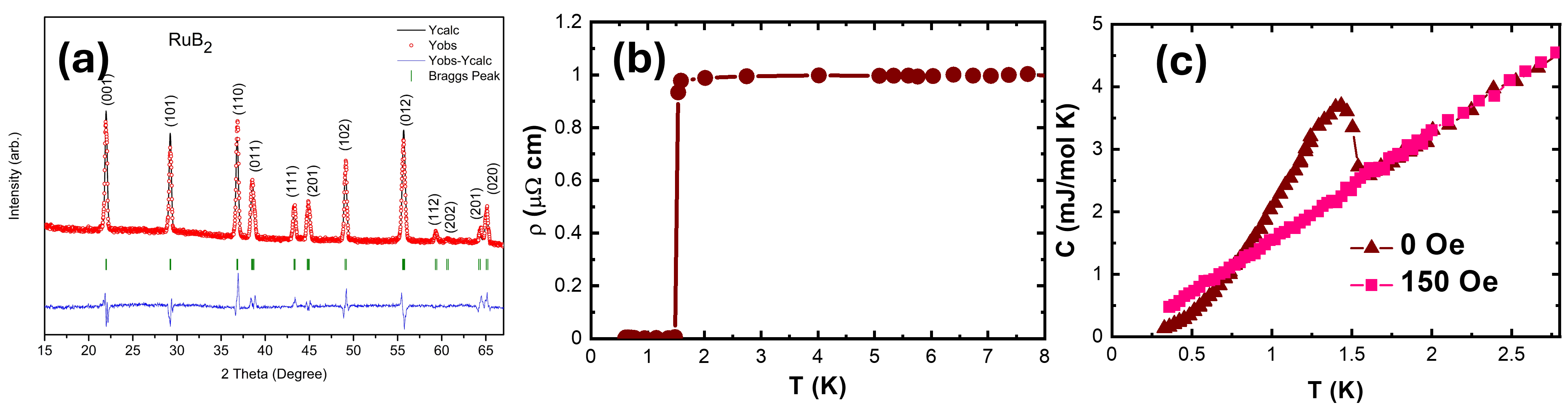} 
\caption{(Color online) (a) The powder xrd data and refinement, (b) the low temperature electrical resistivity, and (c) the heat capacity in zero magnetic field as well as in a field of 150~Oe\@.     
\label{FigS1}}
\end{figure*}

Polycrystalline samples of RuB$_2$ were synthesized by arc-melting stoichiometric ratios of Ru (5N, Alfa Aesar) and B (6N, Alfa Aesar) as described in detail previously \cite{Singh2007,Jaskaran}.  Powder x-ray diffraction shown in Fig. \ref{FigS1}(a) confirmed that the synthesized material is single phase and a refinement of the powder pattern gave lattice parameters which match well with the reported values \cite{Singh2007,Jaskaran}.  The electrical transport and heat capacity down to $350$~mK was measured using the He3 insert in a QD-PPMS and are shown in Fig. \ref{FigS1}(b) and (c), respectively.
\section{$\mu$SR Measurements}  
\subsection{TF-$\mu$SR Measurements}
To probe the superconducting nature of the material RuB$_2$, we applied an external magnetic field 60 G in transverse direction to the initial spin direction of muons. In an applied field, muon spin precess around that field with Larmor Frequency written as:
\[ \nu = \frac{\gamma_\mu}{2\pi} B \]
The time evolution of the muon spin polarization in the applied transverse field was recorded and analyzed over the temperature range 0.01--2~K. Fig.~1(a) in the main text presents the corresponding asymmetry versus time spectra at selected temperatures: 2~K, which is well above $T_c$, and 0.7~K, which is well below $T_c$. Below the superconducting transition temperature ($T_c$), an enhanced damping rate is observed, which is attributed to the formation of a flux-line lattice in the superconducting state. This results in an inhomogeneous magnetic field distribution within the sample, leading to a faster depolarization of the muon signal below $T_c$.\\

In contrast, as above $T_c$, the system is in the normal state, where the magnetic field distribution remains relatively homogeneous. A small relaxation is observed, as evident from the spectra collected at 2~K, arising primarily from static nuclear magnetic moments along with a minor background contribution.\\
The data below $T_c$ are well described by a sum of two oscillatory Gaussian-damped cosine functions: one corresponding to the sample contribution and the other to a non-relaxing background signal. Above $T_c$, the spectra can be described by a single Gaussian-damped cosine function, given by:
\begin{equation}
A_{\text{TF}}(t) = A_{\text{TF}}(0)\,\exp\left(-\frac{\sigma_{\text{tot}}^2 t^2}{2}\right)\cos\left(\gamma_\mu B_{\text{int}} t + \phi\right),
\label{eq:gcosine}
\end{equation}
where \(A_{\text{TF}}(0)\) is the initial asymmetry, \(\gamma_{\mu}/2\pi = 135.5~\text{MHz/T}\) is the muon gyromagnetic ratio, \(B_{\text{int}}\) is the internal magnetic field sensed by the implanted muons, and \(\phi\) is the initial phase offset of the muon spin polarization. Fig.~1(b) in the main text shows the corresponding Fourier data of the signal for both below and above $T_c$, showing a shift in the field sensed by the muons when the sample is in superconducting state. As can be seen from the inset of Fig.~1(b), we observe an anomalous paramagnetic shift of the internal field experienced by the muon in the superconducting state, rather than the conventionally expected diamagnetic shift. A similar effect has been reported in several compounds, including Ta$_2$V$_{3.1}$Si$_{0.9}$ and Nb$_{0.25}$Bi$_2$Se$_3$.

In general, such a paramagnetic shift can arise from several possible mechanisms, including field-induced magnetism~\cite{PhysRevLett.103.067010,PhysRevB.93.094513}, vortex disorder~\cite{PhysRevLett.106.127002}, odd-parity superconducting pairing~\cite{PhysRevLett.125.026802}, suppression of the negative Knight shift below $T_c$ due to singlet pairing, or the presence of paramagnetic regions within the sample~\cite{PhysRevB.102.134514}. In particular, the superconducting regions expel magnetic flux, which may lead to an effective enhancement of the local field in adjacent paramagnetic regions. Further experimental investigations are required to elucidate the origin of the paramagnetic shift observed in RuB$_2$.

The total relaxation rate obtained from the fitting has contribution due to superconducting vortex-lattice (\(\sigma_{\text{sc}}\)) and contribution from the nuclear magnetic moments (\(\sigma_{\text{nm}}\)) and can be calculated as $$\sigma_{\text{tot}}^2 = \sigma_{\text{sc}}^2 + \sigma_{\text{nm}}^2$$. As the contribution due to nuclear magnetic moments is temperature independent and above transition temperature the superconducting contribution is zero, we can calculate $\sigma_{\text{sc}}$ as a function of temperature which is purely due to the superconducting nature of the material and is shown in Fig.~\ref1(c). This relaxation rate due to vortex lattice is directly related to the magnetic penetration depth and the superconducting gap as \cite{tinkham2004introduction,gupta2020isotropic}:
\[\frac{\sigma_{\text{sc}}(T)}{\sigma_{\text{sc}}(0)} =\frac{\lambda_L^{-2}(T)}{\lambda_L^{-2}(0)}\]
\begin{equation}
= 1 + \frac{1}{\pi} \int_{0}^{2\pi} 
\int_{\Delta(T, \varphi)}^{\infty}
 \left(\frac{\partial f}{\partial E}\right) \frac{E dE d\varphi}{\sqrt{E^2 - \Delta(T, \varphi)^2}}
\label{eq:sigma}
\end{equation}\\
where $f = [1 + \exp(E/k_B T)]^{-1}$ is the Fermi function, and \( E = \sqrt{\epsilon^2 + \Delta^2(T, \psi)} \) is the quasiparticle energy. Here, \( \Delta(T, \psi) \) represents the superconducting order parameter, which depends on both the temperature \( T \) and the azimuthal angle \( \psi \). The temperature dependence of the superconducting gap \( \Delta(T, \psi) \) is given by the empirical relation \( \Delta(T, \psi) = \Delta_0 \delta(T/T_c) g(\psi) \), where \( \delta(T/T_c) = \tanh\left( 1.82 \left[ 1.018 \left( \frac{T_c}{T} - 1 \right) \right]^{0.51} \right) \)~\cite{carrington2003magnetic,khasanov2020single}, and \( g(\psi) \) represents the angular dependence of the gap. For the s-wave model, \( g(\psi) = 1 \), whereas for the d-wave model, \( g(\psi) = \sqrt{2} \cos(2\psi) \). The experimental data are well described by an isotropic $s$-wave superconducting gap, yielding a gap value of $\Delta_0 = 0.16 \pm 0.003$~meV. This corresponds to a ratio $\Delta_0/k_B T_c = 1.67 \pm 0.03$, which is close to the BCS weak-coupling limit of 1.76. This result indicates that the superconductivity in the material is consistent with conventional BCS behavior, characterized by a fully gapped, isotropic pairing symmetry and weak electron-phonon coupling.

\subsection{ZF-$\mu$SR Measurements}
Additionally, ZF-$\mu$SR measurements are also performed in a true zero magnetic field environment. The evolution of muon relaxation asymmetry spectra with time is shown in Fig. 2(a) of the main text. The spectra is well described by a damped static Gaussian Kubo-Toyabe (KT) function~\cite{Gupta2021gap}:

\begin{equation}
A(t) = A_s \exp(-\Lambda t) G_{\text{KT}}(t) + A_{bg},
\label{GTfunction}
\end{equation}

where \(A(t)\) is the measured asymmetry as a function of time, \(A_s\) is the sample asymmetry and \(A_{bg}\) is contribution due to background, both are independent of temperature. The term \(\exp(-\Lambda t)\) represents exponential damping due to dynamic electronic spin relaxation, with \(\Lambda\) being the associated relaxation rate. The function \(G_{\text{KT}}(t)\) is the static Gaussian Kubo-Toyabe function, which models muon depolarization arising from randomly oriented, quasi-static local magnetic fields caused primarily by nuclear magnetic moments~\cite{kubo1981stochastic}.\\

The functional form of the Gaussian Kubo-Toyabe \cite{kubo1981stochastic} function is given by:
\begin{equation}
G_{\text{KT}}(t) = \frac{1}{3} + \frac{2}{3} \left( 1 - \sigma_{\text{ZF}}^2 t^2 \right) \exp\left( -\frac{\sigma_{\text{ZF}}^2 t^2}{2} \right),
\end{equation}
 Here, \(\sigma_{\text{ZF}}\) represents the width of the static local field distribution arising from randomly oriented nuclear dipolar moments. The average local magnetic field experienced by the muon spin ensemble is given by \(\langle B_\mu \rangle = \sigma_{\text{ZF}} / \gamma_\mu\), where \(\gamma_\mu = 135.53\ \text{MHz/T}\) is the muon gyromagnetic ratio. In our analysis, we considered both cases by allowing \(\sigma_{\mathrm{ZF}}\) and \(\lambda_{\mathrm{ZF}}\) to vary independently, as well as by fixing \(\lambda_{\mathrm{ZF}}\) to the an average value of 0.074 $\mu\mathrm{s}^{-1}$. In both analyses, \(\sigma_{\mathrm{ZF}}\) exhibits an evident increase below \(T_c\). For reference, the ZF spectra collected at the DOLLY spectrometer is shown in Fig. S2 (a) and temperature dependence of relaxation $\sigma(T)$ is shown in fig. S2(b).

\begin{figure}[t]
    \centering
    \includegraphics[width=1.0\linewidth]{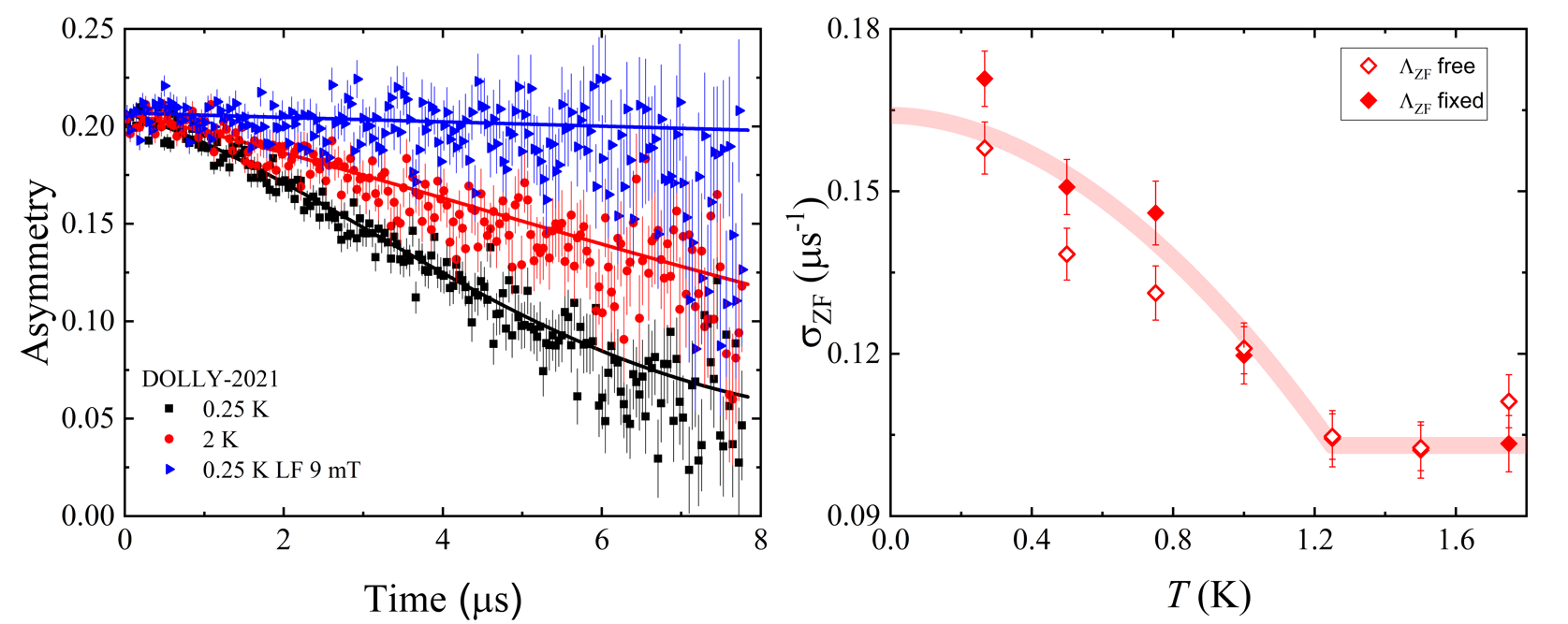}
    \caption{(a) ZF-$\mu$SR spectra of RuB$_2$ measured on the DOLLY spectrometer below and above the superconducting transition temperature. The application of a small longitudinal field of 9~mT decouples the muon spins from the weak internal fields. (b) Temperature dependence of the zero-field muon relaxation rate, $\sigma_{\mathrm{ZF}}$, showing a clear increase below $T_c$, indicative of time-reversal symmetry breaking (TRSB).}
    \label{fig:placeholder}
\end{figure}

\begin{figure}[t]
    \centering
    \includegraphics[width=1.0\columnwidth]{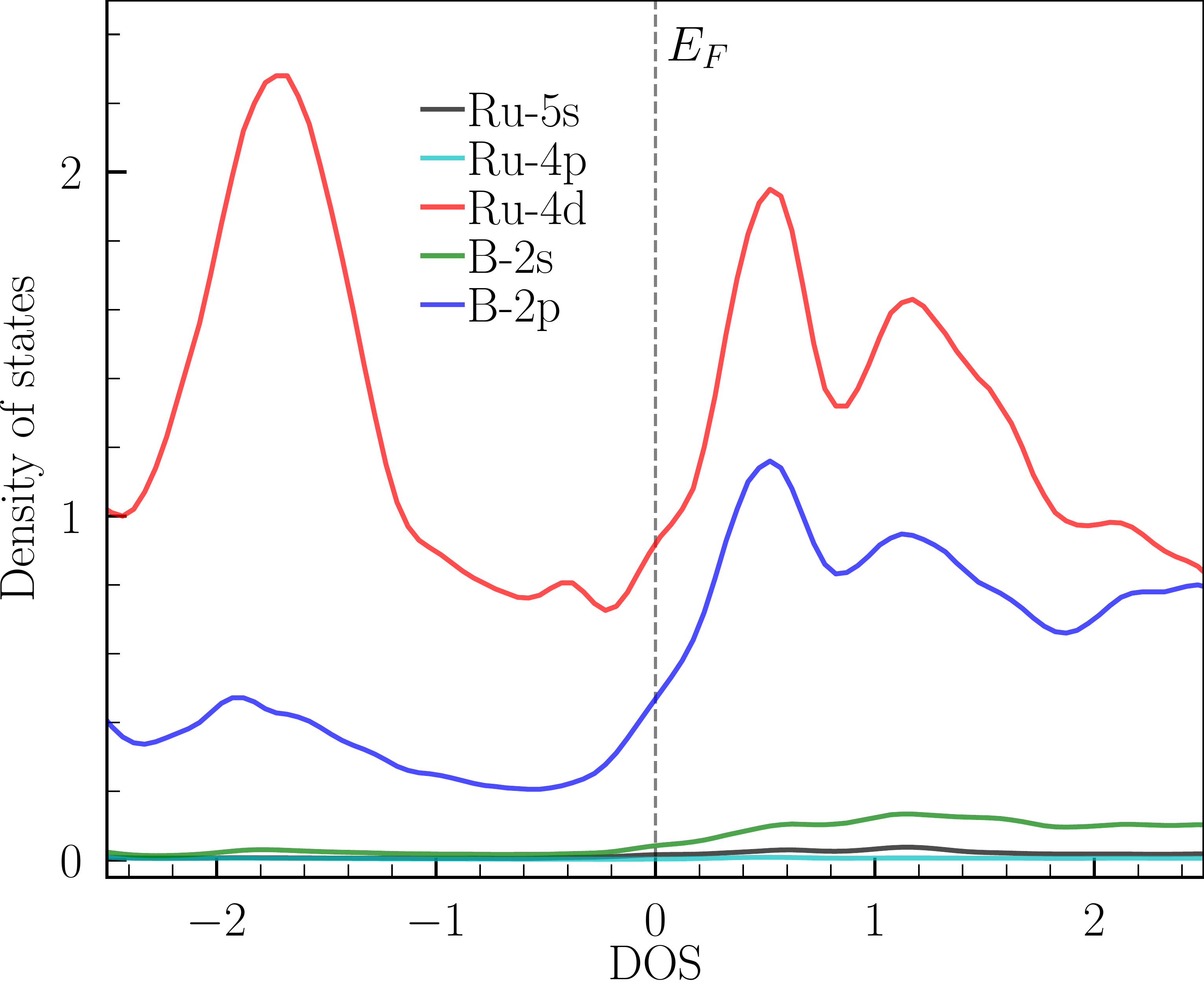}
    \caption{Orbital-resolved density of states of RuB$_2$ calculated without spin-orbit coupling.}
    \label{fig:pdos}
\end{figure}

\section{Contribution to density of states at Fermi level from each Fermi sheet}

Figure~S3 shows the orbital-resolved projected density of states (PDOS) of RuB$_2$ calculated in the absence of spin–orbit coupling (SOC). The low-energy electronic structure near the Fermi level is predominantly derived from Ru-$4d$ orbitals, with additional substantial contributions from B-$2p$ states and only a minor contribution from B-$2s$ orbitals. The pronounced Ru-$4d$ character around the Fermi energy naturally accounts for the appreciable SOC-induced band splitting observed in RuB$_2$.\\

\begin{figure}[h]
    \centering
    \includegraphics[width=1.0\columnwidth]{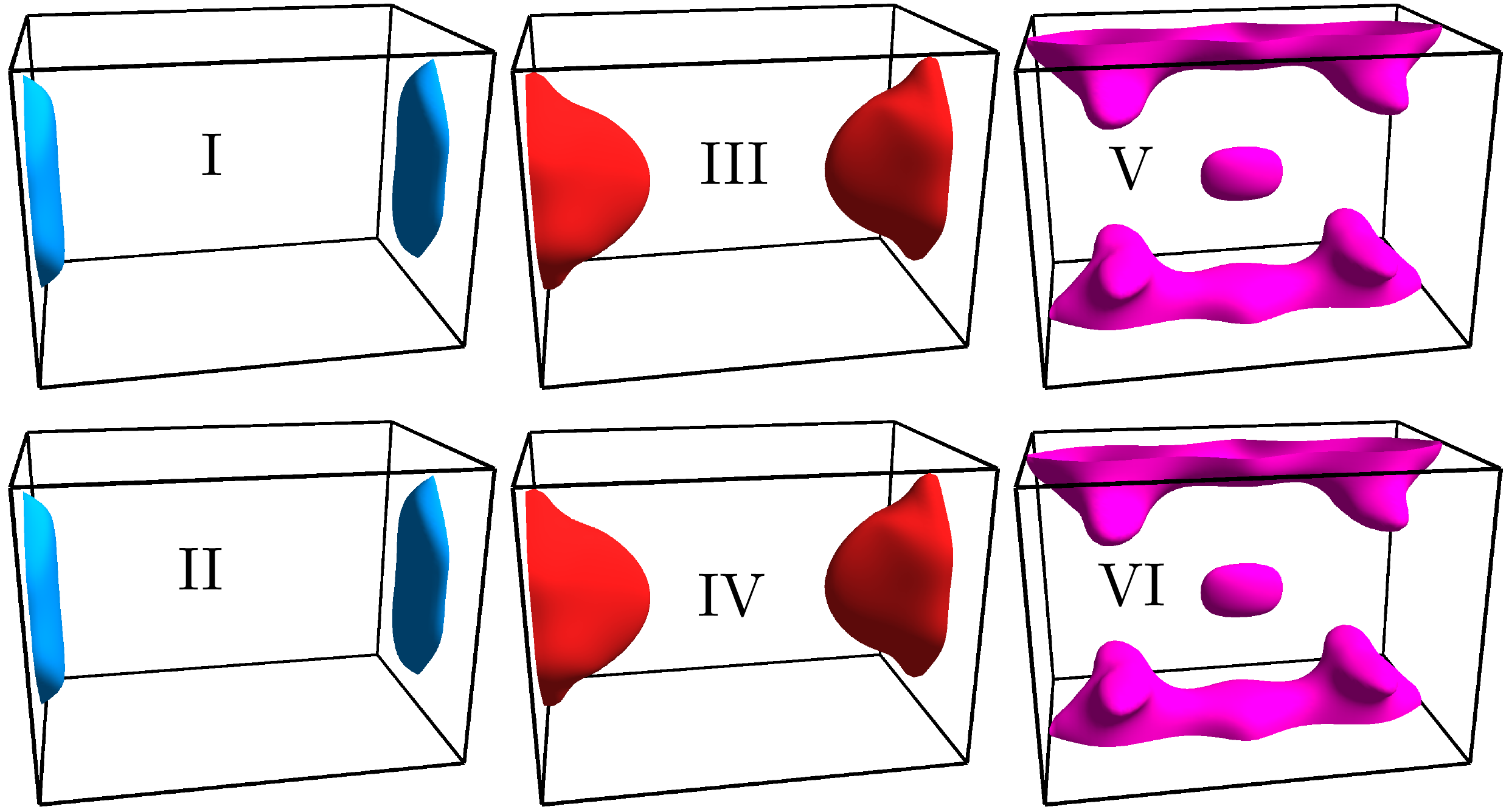}
    \caption{Representative Fermi sheets of RuB$_2$ including the effect of SOC.}
    \label{fig:fermi_sheets}
\end{figure}

\begin{figure}[h]
    \centering
    \includegraphics[width=1.0\columnwidth]{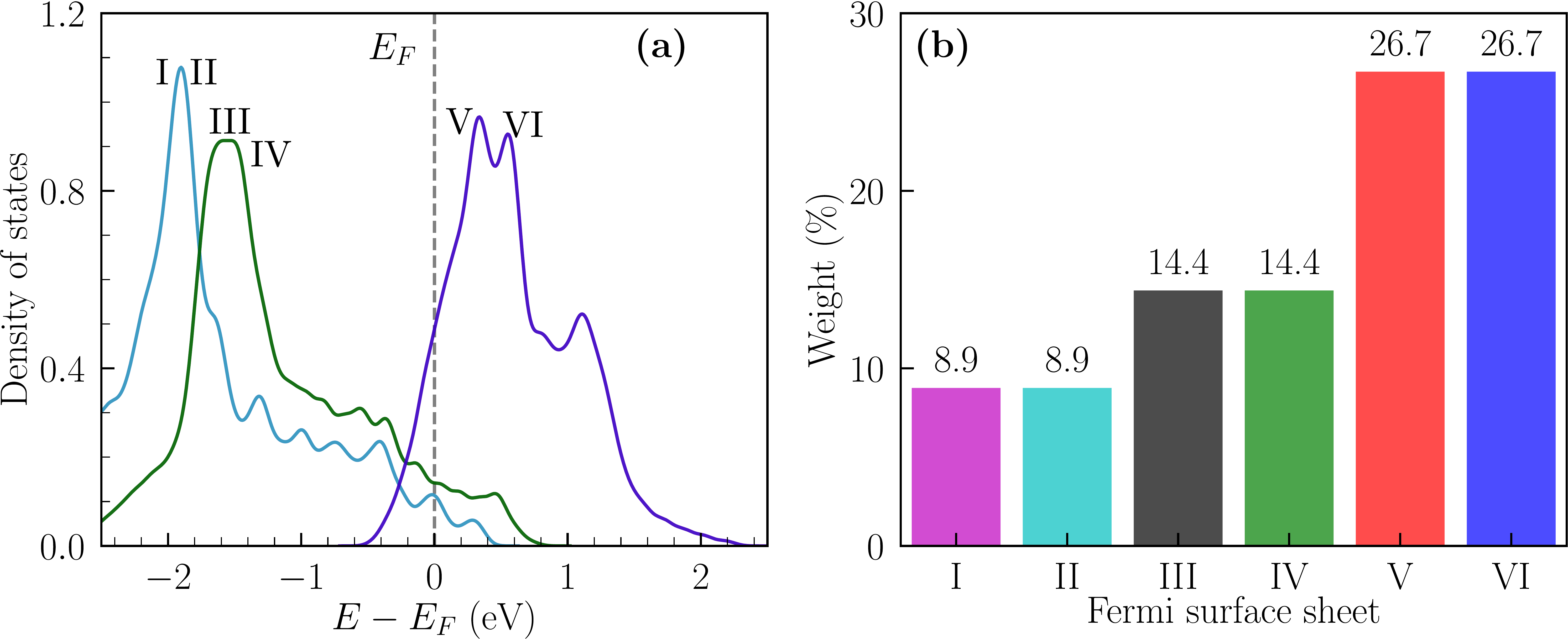}
    \caption{(a) The contribution to the DOS of the individual Fermi surfaces shown in Fig. S3. (b) The relative weights of the DOS contributions of each Fermi surface to their total DOS at $E_F$.}
    \label{fig:fermi_sheets_weight}
\end{figure}

To further investigate the multiband electronic structure, we constructed a tight-binding Hamiltonian using maximally localized Wannier functions (MLWFs) and computed the Fermi surfaces including SOC. As shown in Fig.~S4, RuB$_2$ hosts six distinct Fermi surface sheets, each corresponding to a band crossing the Fermi level. The relative contribution of each sheet to the density of states at the Fermi level is presented in Fig.~S5(a). Notably, Fermi sheets V and VI together account for nearly half of the total density of states at the Fermi level as illustrated in Fig.~S5(b), highlighting the pronounced multiband character of the system.\\
\bibliography{RuB2_refs}

\end{document}